# Coexistence of superconductivity and antiferromagentic order in $Er_2O_2Bi$ with anti-$ThCr_2Si_2$ structure

Lei Qiao[1], Ning-hua Wu[1], Tianhao Li[1], Siqi Wu[1], Zhuyi Zhang[1], Miaocong Li[1], Jiang Ma[1], Baijiang Lv[2], Yupeng Li[1], Chenchao Xu[1], Qian Tao[1], Chao Cao[3], Guang-Han Cao[1,4], and Zhu-An Xu[1,4*]

[1] *Zhejiang Province Key Laboratory of Quantum Technology and DeviceDepartment of Physics, Zhejiang University, Hangzhou 310027, China;*
[2] *Key Laboratory of Neutron Physics and Institute of Nuclear Physics and Chemistry, China Academy of Engineering Physics, Mianyang 621999, China;*
[3] *Department of Physics, Hangzhou Normal University, Hangzhou 310036, China; and*
[4] *State Key Laboratory of Silicon Materials, Zhejiang University, Hangzhou 310027, China;*

(Dated: May 18, 2021)

We investigated the coexistence of superconductivity and antiferromagnetic order in the compound $Er_2O_2Bi$ with anti-$ThCr_2Si_2$-type structure through resistivity, magnetization, specific heat measurements and first-principle calculations. The superconducting transition temperature $T_c$ of 1.23 K and antiferromagnetic transition temperature $T_N$ of 3 K are observed in the sample with the best nominal composition. The superconducting upper critical field $H_{c2}(0)$ and electron-phonon coupling constant $\lambda_{e-ph}$ in $Er_2O_2Bi$ are similar to those in the previously reported non-magnetic superconductor $Y_2O_2Bi$ with the same structure, indicating that the superconductivity in $Er_2O_2Bi$ may have the same origin as in $Y_2O_2Bi$. The first-principle calculations of $Er_2O_2Bi$ show that the Fermi surface is mainly composed of the Bi $6p$ orbitals both in the paramagnetic and antiferromagnetic state, implying minor effect of the $4f$ electrons on the Fermi surface. Besides, upon increasing the oxygen incorporation in $Er_2O_xBi$, $T_c$ increases from 1 to 1.23 K and $T_N$ decreases slightly from 3 to 2.96 K, revealing that superconductivity and antiferromagnetic order may compete with each other. The Hall effect measurements indicate that hole-type carrier density indeed increases with increasing oxygen content, which may account for the variations of $T_c$ and $T_N$ with different oxygen content.

## I. INTRODUCTION

The interplay of superconductivity and magnetism has been a research focus for decades. Usually, superconductivity and magnetic order will compete with each other, and thus superconductivity seldom occurs in strong magnetic materials [1]. In conventional s-wave superconductors, local magnetic moments would break up spin singlet Cooper pairs and hence strongly suppress superconductivity [2]. In the case of rare-earth superconductors, $4f$ electrons may have a localized magnetic moment which could weakly couple with conduction electrons, and the coexistence of superconductivity and magnetism is possible to realize. There have been some examples of such materials, like $RMo_6S_8$ [3, 4], $RRh_4B_4$ [3, 4], $RBa_2Cu_3O_{7-\delta}$ [5–8], $RNi_2B_2C$ [9, 10], $RPdBi$ [11] ($R$ = some of the rare earth elements), and iron-based superconductors containing rare earth elements [12]. In ferromagnetic $ErRh_4B_4$ [13] and $HoMo_6S_8$ [14], and antiferromagntic (AFM) $GdMo_6S_8$[15], the competition between superconductivity and magnetic long-range order has resulted in the reentrant superconductivity behavior. In the $RNi_2B_2C$ and $RPdBi$ systems, the competitive relationship between AFM transition temperature $T_N$ and superconducting (SC) transition temperature $T_c$ can be described by the Gennes scaling[10, 11]. Moreover, the coexistence of magnetism and superconductivity may lead to some exotic phases. For example, $UTe_2$ where superconductivity coexists with ferromagnetic spin fluctuations is proposed to have a spin-triplet pairing state [16], and the coexistence of magnetism and superconductivity in the $RPdBi$ systems can serve as a unique platform to investigate topological orders with multisymmetry breaking states[11].

The $R_2O_2Bi$ ($R$ = rare earth) family are a series of layered compounds with anti-$ThCr_2Si_2$ structure. Superconductivity has been found by excess oxygen incorporation in some of $R_2O_2Bi$ members, irrespective of the presence of magnetic ordering [17]. Superconductivity in nonmagnetic $Y_2O_2Bi$ was found dependent on the oxygen content[18]. The oxygen incorporation will cause the expansion of inter-Bi-layer distance, which is proposed to be crucial to the occurrence of superconductivity. The obvious change in the carrier density companied with oxygen incorporation was not found. Meanwhile, the search for superconductivity in the magnetic $R_2O_2Bi$ system has produced various outcomes. The $Ce_2O_2Bi$ was found to be an AFM Kondo lattice compound without observation of superconductivity for temperature down to 0.3 K [19]. Although $Er_2O_2Bi$, $Tb_2O_2Bi$ and $Dy_2O_2Bi$ have been reported to be superconducting, the detailed investigation of their properties is lacked[17, 20]. Whether the $4f$ electron magnetism has remarkable influence on superconductivity is an interesting issue and it is worthy of careful study. Up to now, among the magnetic $R_2O_2Bi$ members, only the $Er_2O_2Bi$ is found to be superconducting without the CaO incorporation which serves as an oxidant to obtain excess oxygen in $R_2O_2Bi$ [17].

However, the Ca ion impurity would also be incorporated in the samples and it may affect the superconducting properties. Thus it is better to tune the oxygen content without CaO incorporation in studying the interplay of magnetism and superconductivity in a series of $R_2O_2Bi$.

Here, we systematically studied the $Er_2O_2Bi$ polycrystalline compounds with various oxygen content by performing magnetization, specific heat and resistivity measurements. The antiferromagnetic (AFM) transition associated with the $Er_2^{3+}O_2^{2-}$ layer occurs at $T_N$ of about 3 K, which indicates that there is a coexistence of superconductivity and AFM order in $Er_2O_2Bi$. The first-principle calculations suggest that the Fermi surface of $Er_2O_2Bi$ is similar to that of $Y_2O_2Bi$, implying superconductivity in both compounds could share the same mechanism and the conducting monatomic $Bi^{2-}$ square net layer may be crucial to the occurrence of superconductivity. Above $T_N$, there seems a kind of short-range magnetic correlation which enhances the specific heat significantly and reduces the magnetic entropy above $T_N$. Furthermore, we have synthesized four samples with different nominal oxygen content $x$ in the $Er_2O_xBi$ compounds ($x$ = 1.8, 1.9, 2.0 and 2.1). Upon increasing $x$, the actual oxygen content indeed increases, and $T_N$ is slightly suppressed. Meanwhile the hole-type carrier density and $T_c$ both increases, indicating that the AFM order may compete with superconductivity. The variation of hole-type carrier density may account for the change of $T_c$ and $T_N$. This work provides a new example to study the interplay of rare earth magnetism and superconductivity in such an anti-$ThCr_2Si_2$-type structure.

## II. EXPERIMENTAL DETAILS

The $Er_2O_xBi$ ($x$ = 1.8, 1.9, 2.0, 2.1) polycrystalline compounds were synthesized by solid-state reactions. High-purity $Er_2O_3$ (Alfa, 99.99%), Er (Alfa 99.8%) and Bi (Alfa 99.999%) were used as starting materials. The mixed powders were pressed into pellets and covered with Ta foils. An excess amount of Bi was added because of its high vapor pressure. The pellets was sealed in an evaluated silica tube, and heated at 1000 °C for 20 h in furnace. After it had cooled down to room temperature naturally, the pellets was reground and pressed again, then annealed at 1000 °C to improve the sample homogeneity. All the operations were taken in an Ar-filled glove box except the heating processes.

The powder x-ray diffraction (XRD) patterns were recorded on a PANalytical X-ray diffractometer with Cu K$\alpha$ radiation at room temperature. We used the Rietveld refinements employing program GSAS with EXPGUI[21] to obtain lattice parameters. The DC magnetization was measured on two magnetic property measurement systems (MPMS-XL5 for $T$ > 1.8 K and MPMS3 with $^3$He refrigerator for $T$ < 1.8 K). Electrical resistivity and Hall effect were measured by a standard four-probe method on a physical property measurement system(PPMS-9). Specific heat measurements were also carried out on the same PPMS-9 by a heat-pulse relaxation method. We also used an Oxford superconducting magnet system equipped with a $^3$He cryostat to measure the superconducting transitions.

The first-principle calculations were performed with the Vienna Ab Initio Simulation Package (VASP) code [22, 23]. Throughout the calculations, the Perdew, Burke, and Ernzerhof (PBE) parameterization of generalized gradient approximation (GGA) to the exchange correlation function was used [24]. The spin-orbit coupling (SOC) is included using the second variational method. The energy cutoff of plane-wave basis was chosen to be 600 eV, which was sufficient to converge the total energy to 1 meV/atom. A $\Gamma$-centered 18 × 18 × 18 Monkhorst-Pack [25] k-point mesh was chosen to sample the Brillouin zone in the calculations. Considering that the Er-4$f$ states are fully localized at high temperature, the paramagnetic (PM) band structure was calculated by assuming the Er-4$f$ open-core electron. While for the AFM state, the states of $f$ electrons were treated as semi-core valence states by performing LDA+U with U = 6 eV [26–29].

## III. RESULTS AND DISCUSSION

Figure 1(a) shows the XRD patterns of $Er_2O_xBi$ with different nominal oxygen content ($x$). It is obvious that all the four samples have the same main phase of anti-$ThCr_2Si_2$ type structure with a few minor impurity phases depending on the oxygen content. From the XRD patterns, the sample with $x$ = 2.0 has the best quality with least amount of impurity phases, therefore the most of superconducting property' characterization is performed based on this sample. We used the Rietveld refinements employing program GSAS with EXPGUI [21] to obtain lattice parameters of the four samples. Figure 1(b) shows the Rietveld refinement of the XRD pattern of the sample $Er_2O_2Bi$. All diffraction peaks can be indexed with the crystalline structure of $Er_2O_2Bi$, and their ($h\ k\ l$) indices are also indicated for each peak. The fitting parameters are $a$ = 3.8459(1) Å and $c$ = 13.1585(5) Å with the parameters $R_{wp}$ = 7.65% and $\chi^2$ = 1.93 which indicate a good convergence. The inset of Figure 1(a) displays the variations of the lattice constants $a$ and $c$ with nominal oxygen content. As the oxygen content $x$ increases, there is a remarkable increase in the lattice constant $c$, and the lattice constant $a$ remains almost constant. This $c$-axis expansion due to excess oxygen incorporation is also observed in $Y_2O_2Bi$ [18]. The inset of Figure 1(b) displays the anti-$ThCr_2Si_2$-type structure of $Er_2O_2Bi$ which is composed of a Bi square net layer and two ErO layers.

Figure 2(a) shows the temperature-dependent resistiv-ity for the sample $Er_2O_2Bi$ from 300 to 0.4 K. From the inset of Figure 2(a), there are two obvious transitions in resistivity which should correspond to $T_c$ and $T_N$, respectively. The $\chi-T$ curves around $T_N$ are also plotted in the inset for comparison, which indicates the good coincidence between the transitions in both resistivity and magnetic susceptibility at $T_N$. The increased magnetic field does not depress the transition in magnetic susceptibility around $T$ of 3 K, excluding a possible superconducting transition. Figure 2(b) shows the resistivity of $Er_2O_2Bi$ near the $T_c$ under different magnetic fields. The resistivity decreases to zero abruptly below the SC transition temperature $T_c = 1.23$ K under zero magnetic filed. Here we define $T_c$ as the temperature at which the resistivity drops to 50% of normal-state resistivity just above the SC transition. The $T_c^{\text{zero}}$ (zero resistivity criterion) is 1.17 K. With the magnetic field increasing, $T_c$ is suppressed continuously. The reduced temperature ($T/T_c$) dependence of upper critical magnetic field $H_{c2}$ is presented in Figure 2(c). We tried to use the one-band Ginzburg-Landau theory, $H_{c2} = H_{c2}(0)(1-t^2)/(1+t^2)$ (where $t = T/T_c$), to fit the upper critical field and the fitting result is not good. It is found that there seems to be a positive curvature close to $T_c$, which could be a characteristic of multi-band superconductor [30–33]. Therefore, we introduce the two-band Ginzburg-Landau theory [34],

$$a_0[\ln t + U(h/t)][\ln t + U(\eta h/t)] + a_1[\ln t + U(h/t)] + a_2[\ln t + U(\eta h/t)] = 0 \quad (1)$$

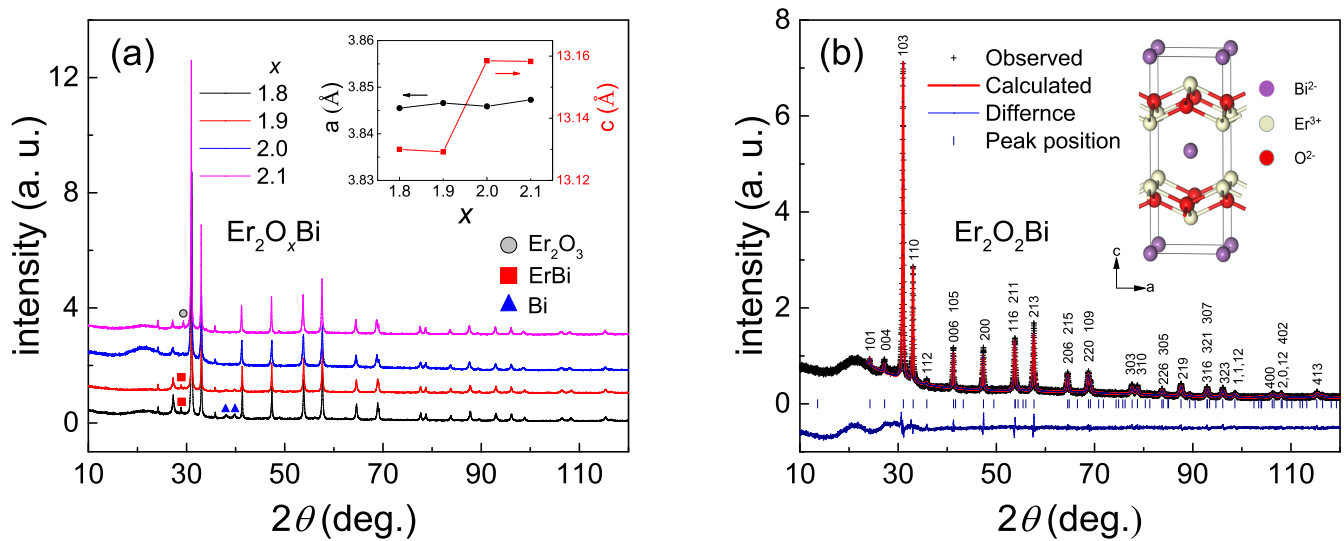

FIG. 1. (Color online) (a) XRD patterns of $Er_2O_xBi$ with different value of $x$. The grey circle, the red rectangle, the blue triangle points represent the peak of impurity $Er_2O_3$, ErBi and Bi, respectively. The inset shows the lattice constants $a$ and $c$ of these four samples. (b) Rietveld refinement of the XRD pattern of the sample $Er_2O_2Bi$. Inset shows the anti-$ThCr_2Si_2$-type structure of $Er_2O_2Bi$.

. Here $a_0 = 2(\lambda_{11}\lambda_{22}-\lambda_{12}\lambda_{21})$, $a_1 = 1+(\lambda_{11}-\lambda_{22})/\lambda_0$, $a_2 = 1-(\lambda_{11}-\lambda_{22})/\lambda_0$, $\lambda_0 = [(\lambda_{11}-\lambda_{22})^2+4\lambda_{12}\lambda_{21}]^{\frac{1}{2}}$, $h = H_{c2}D_1/2\phi_0T$, $\eta = D_1/D_2$, $U(x) = \Psi(x+1/2)\text{-}\Psi(x)$, $\Psi(x)$ is the digamma function. $\lambda_{11}$ and $\lambda_{22}$ are the intraband BCS coupling constants, $\lambda_{12}$ and $\lambda_{21}$ are the interband BCS coupling constants, and $D_1$ and $D_2$ are the in-plane diffusivity of each band. The fitting parameters are listed in Table 1.

Table 1  The fit parameters of two-band model.

| $\lambda_{11}$ | $\lambda_{22}$ | $\lambda_{12}$ | $\lambda_{21}$ | $\eta$ | $H_{c2}(0)$(T) |
|---|---|---|---|---|---|
| 0.95 | 0.51 | 0.29 | 0.84 | 9 | 0.061 |

Based on this two-band model, we can obtain $H_{c2}(0)$ = 0.061 T which is lower than the BCS Pauli limit, i.e., $\mu_0 H_P^{BCS}(0) = 1.84 T_c \sim 2.3$ T. Compared to the upper critical field of $Y_2O_2Bi$ which is about 0.0574 T [18], $H_{c2}(0)$ of $Er_2O_2Bi$ does not show much difference, in spite of the substitution of non-magnetic $Y^{3+}$ ion by the magnetic $Er^{3+}$ ion. It indicates that superconductivity in both compounds could share the same mechanism and the conducting monatomic $Bi^{2-}$ square net layer may be crucial to the occurrence of superconductivity. Then, we estimate Ginzburg-Landau coherence length by $H_{c2} = \Phi_0/2\pi\xi^2$ and $\xi(0)$ of about 730 Å is obtained. Apparently $\xi(0)$ is much larger than the lattice constants, while the length of magnetic unit cell is usually comparable to the lattice constants. The larger $\xi(0)$ value (than the length of magnetic unit cell) could account for the coexistence of superconductivity and antiferromagnetism [10].

Figure 2(d) shows the temperature dependence of magnetic susceptibility $\chi$ (left axis) and inverse magnetic sus-ceptibility, $1/\chi$ (right axis) measured under the magnetic field $H = 100$ Oe. The $\chi(T)$ satisfies the modified Curie-Weiss law $\chi = \chi_0 + \mathcal{C}/(T - \theta_W)$ where $\chi_0$ is a temperature independent susceptibility, $\mathcal{C}$ is the Curie constant and $\theta_W$ is the Weiss temperature. The red solid line shows the Curie-Weiss fit in the temperature range between 7 - 300 K and the derived parameters $\mathcal{C} = 11.77$ emu/mol-Er K, $\chi_0 = 0.00372$ emu/mol-Er and $\theta_W = -22$ K. The negative Weiss temperature indicates an AFM exchange interaction between the magnetic moments. The calculated effective magnetic moment based on the Curie constant is 9.7 $\mu_B$/Er which is very close to that of the free $Er^{3+}$ ion (9.8 $\mu_B$/Er). It implies that the $4f$ electrons of $Er^{3+}$ ion are very localized in the temperature range between 7 - 300 K. Figure 2(e) displays the temperature dependence of magnetic susceptibility $\chi$ from $T = 0.4$ to 5 K. It is clear that there exist two transitions which correspond to the AFM order and SC transition respectively. The red hollow circles and black solid circles stand for the field cooling (FC) and zero-field cooling (ZFC) processes under the magnetic field $H = 10$ Oe around $T_c$, respectively. The data represented by the blue solid squares is measured under $H = 100$ Oe to show the change around $T_N$. To avoid demagnetization effect, the sample is shaped into a long and narrow cuboid (about $3 \times 0.66 \times 0.4$ mm$^3$) and the magnetic field is applied along the long axis of the sample. The SC volume fraction is estimated to be about 133% based on the ZFC data, confirming the bulk nature of superconductivity in $Er_2O_2Bi$. Although the demagnetization factor is close to 0, the overestimated SC volume fraction which is beyond 100% may be caused by the error in measuring the sample volume. The inset of Figure 2(e) is the field dependence of magnetization $M(H)$ at $T = 0.5$ K which is a typical behavior of type II superconductor. Figure 2(f) shows the magnetic susceptibility $\chi$ measured in the temperature $T$-sweep mode containing both $T$-up and $T$-down processes in the temperature range between 2 - 4 K. To avoid the error of $T$, the speed of $T$ sweep is as slow as 0.15 K/min. There is a clear magnetic hystere-sis around $T$ of 3 K under 100 Oe, suggesting a possible first-order transition.

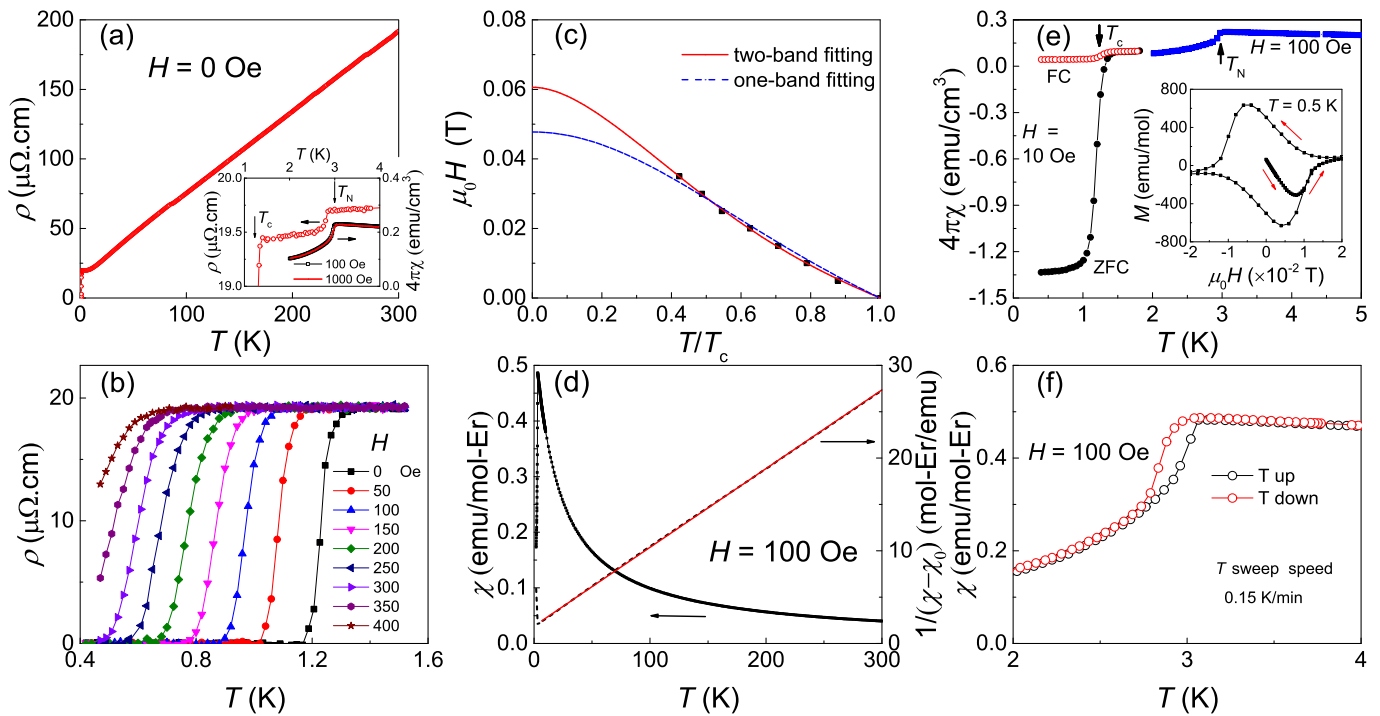

FIG. 2. (Color online) (a) Resistivity of $Er_2O_2Bi$ from 300 to 0.4 K. Inset: temperature dependence of resistivity around two transition temperatures and the temperature dependence of magnetic susceptibility under the magnetic field $H = 100$ and 1000 Oe around $T_N$. (b) Resistivity of $Er_2O_2Bi$ near the $T_c$ under different magnetic fields. (c) Normalized temperature dependence of upper critical field of $Er_2O_2Bi$. The blue dotted line and the red solid line represent the fitting of one-band and two-band Ginzburg-Landau theory, respectively. (d) Temperature dependence of magnetic susceptibility, $\chi$ (left axis) and inverse magnetic susceptibility, $1/\chi$ (right axis) measured under the magnetic field $H = 100$ Oe. The red solid line shows the Curie-Weiss fit over the temperature range 7 - 300 K. (e) Temperature dependence of magnetic susceptibility $\chi$ from $T = 0.4$ to 1.8 K in the zero-field cooling (ZFC) process (black solid circles) and field cooling (FC) process (red hollow circles) under the magnetic field $H = 10$ Oe, and temperature dependence of magnetic susceptibility $\chi$ from $T = 2$ to 5 K under the magnetic field $H = 100$ Oe (blue solid squares). Inset: field dependence of magnetization $M(H)$ at $T = 0.5$ K. (f) Magnetic susceptibility $\chi$ measured in the temperature $T$-sweep mode containing both $T$-up and $T$-down processes between temperature 2 - 4 K.

Figure 3(a) shows the temperature dependence of specific heat $C$ of $Er_2O_2Bi$. The peak at $T_N$ is so sharp, indicating a possible first-order transition. Such a sharp peak in specific heat has been observed in the magnetic first order phase transitions in other compounds, i.e., $DyCu_2Si_2$ [35]. With the magnetic field increasing, the peak is suppressed gradually, consistent with the AFM order. As shown in the inset of Figure 3(a), $T_N$ decreases monotonously with increasing magnetic field. Figure 3(b) displays the temperature dependence of specific heat $C$ of $Er_2O_2Bi$ among the temperature range 0 - 2 K under magnetic fields $\mu_0H = 0$, 0.01, 0.05 and 0.1 T, respectively. There is no obvious peak of specific heat observed at the SC transition temperature $T_c = 1.23$ K, which is probably hidden in the shoulder of the large AFM peak. However, from the inset of Figure 3(b), the different behaviors of $C/T$ at the lowest temperature may be induced by SC transition. The $C/T$ under zero magnetic field has a drop around 0.6 K, while the drop of $C/T$ disappeared under $\mu_0H = 0.01$ T, and $C/T$ become coincident under $\mu_0H = 0.05$ and 0.1 T. Concerning the $C/T$ behavior in $Y_2O_2Bi$ [18], there is also an upturn below about 0.6 K which is proposed to be caused by a Schottky anomaly originating from the $^{209}Bi$ nuclei (I = 9/2). Therefore, we may ascribe the upturn below 0.6 K in $C/T$ of $Er_2O_2Bi$ to the Bi nuclear contributions and the drop below 0.6 K under 0 T to the SC transition. With increasing magnetic field, the SC transition is suppressed and the upturn behavior gradually appears. When the magnetic field is beyond $H_{c2}(T)$, i.e., $\mu_0H = 0.05$ and 0.1 T, the drop in $C/T$ induced by the SC transition disappears and the $C/T$ would remain almost the same with further increasing magnetic field.

Figure 3(c) shows the specific heat divided by temperature, $C/T$, versus $T^2$ under the magnetic field $\mu_0H = 0$ and 9 T. The following equation is used to fit the $C/T$ data,

$$C/T = \gamma + \beta T^2, \qquad (2)$$

where $\gamma$ and $\beta$ denote the coefficient of electron and phonon contributions for $Er_2O_2Bi$, respectively. The derived parameters are $\gamma = 324$ (425) mJ/mol K$^2$ and $\beta = 3.1$ (3.7) mJ/mol K$^4$ under the magnetic field $\mu_0H = 0$ (9) T. The derived Debye temperatures $\Theta_D$ is 146 K and 138 K under $\mu_0H = 0$ and 9 T respectively, by using the formula $\Theta_D = (12\pi^4 NR/5\beta)^{1/3}$. Since the magnetic field usually should not change the phonon contribution to specific heat, the very close values of Debye temperatures $\Theta_D$ for $\mu_0H = 0$ T and 9 T are reasonable. The Debye temperature of $Er_2O_2Bi$ (146 K) is a little smaller than that of $La_2O_2Bi$, which is about 150 K [19]. The difference between their Debye temperatures may result from a little larger mass of Er ions compared to the La ions. Based on $\Theta_D = 146$ K, the electron-phonon coupling constant $\lambda_{e-ph}$ is estimated to be 0.53 from the McMillan equation[36]:

$$\lambda_{e-ph} = \frac{\mu^* \ln(\frac{\Theta_D}{1.45T_c}) + 1.04}{\ln(\frac{\Theta_D}{1.45T_c})(1 - 0.62\mu^*) - 1.04} \qquad (3)$$

where $\mu^*$ is the Coulomb pseudopotential, and an empirical value of 0.13 is used. This value of $\lambda_{e-ph}$ in $Er_2O_2Bi$ is as the same as the value $\lambda_{e-ph} = 0.53$ in $Y_2O_2Bi$, both pointing to the typical weak coupling conventional BCS superconductors.

The Sommerfeld coefficient $\gamma \sim 324$ mJ/mol K$^2$ (under zero magnetic field) is quite large. Usually, the large $\gamma$ value may result from Kondo interactions or strong magnetic fluctuations, which is difficult to distinguish them only based on the specific heat data. After carefully investigating the behaviors of resistivity and magnetic susceptibility, we argue that it should be caused by the magnetic fluctuations rather than the Kondo interactions. There is no obvious feature of Kondo effect in the temperature dependence of resistivity, such as a negative logarithmic temperature dependence of resistivity induced by Kondo scattering. The magnetic moments of $Er^{3+}$ are localized from 300 to 7 K derived from magnetic susceptibility, which is not consistent with the Kondo effect model (magnetic moments will be partially screened by the conduction electrons at low temperature). Instead, it seems that the magnetic fluctuations could account for the large $\gamma$, which is supported by the calculated magnetic entropy (to be shown below). Considering the contribution to specific heat from spin fluctuations in $Er_2O_2Bi$ do not include a $T^3 \ln T$ term which is a character of ferromagnetic fluctuations, the spin fluctuations here may be antiferromagnetic type [37].

Figure 3(d) shows the magnetic entropy calculated by integrating $C_e/T$ over temperature. The $C_e$ denotes the electronic contribution to specific heat, which is separated from the phonon contribution. The phonon contributions are calculated by the Debye model with $\Theta_D = 146$ K. The magnetic entropy does not recover $R\ln2$ at $T_N$ and keeps increasing, and finally saturates as temperature is close to 18 K. Assuming that the ground state of $Er_2O_2Bi$ is Kramers two-fold degenerate, the free magnetic entropy should be $R\ln2$. The reduced magnetic entropy at $T_N$ may be interpreted by a kind of short-range magnetic correlation above the AFM long-range order temperature ($T_N$). The saturation value of magnetic entropy at $T = 18$ K is a little smaller than $R\ln2$, which is probably introduced by the error in calculating the phonon contributions based on the Debye model.

Figure 4(a) shows the temperature dependence of Hall resistivity for the sample $Er_2O_2Bi$. The magnetic field dependence of Hall resistivity at high temperature is nearly linear which can be fitted by the usual one-band model. However, when the temperature decreases, the magnetic field dependence becomes nonlinear, which could be induced by magnetic skew-scattering and the effects of two types of charge carriers. Figure 4(b) displays the temperature dependence of Hall coefficient of $Er_2O_xBi$ with $x$ = 1.8, 1.9 and 2.0, respectively. With increasing nominal oxygen content ($x$), the positive Hall coefficient is decreasing, implying the increase in the hole-type carrier density. Since the samples with $x$ = 2.1 and $x$ = 2.0 have almost the same parameters of $c$-axis, and meanwhile their $T_c$ and $T_N$ are also the same, we believe that both sample should have the same oxygen content, and the excess oxygen content in the sample with nominal $x$ = 2.1 may lead to the formation of a few impurity phases like $Er_2O_3$. Thus the sample with $x$ = 2.1 does not have good quality and its Hall coefficient is not represented for discussions. Figure 4(c) shows the $x$ dependence of carrier density $n_H$ and mobility $\mu$ calculated based the Hall coefficient and resistivity at 100 K. The carrier density is calculated by a single band approximation. It is obvious that there is an increase in the carrier density as $x$ increases from 1.8 to 2.0, which means the oxygen incorporation indeed induce more hole-type charge carriers.

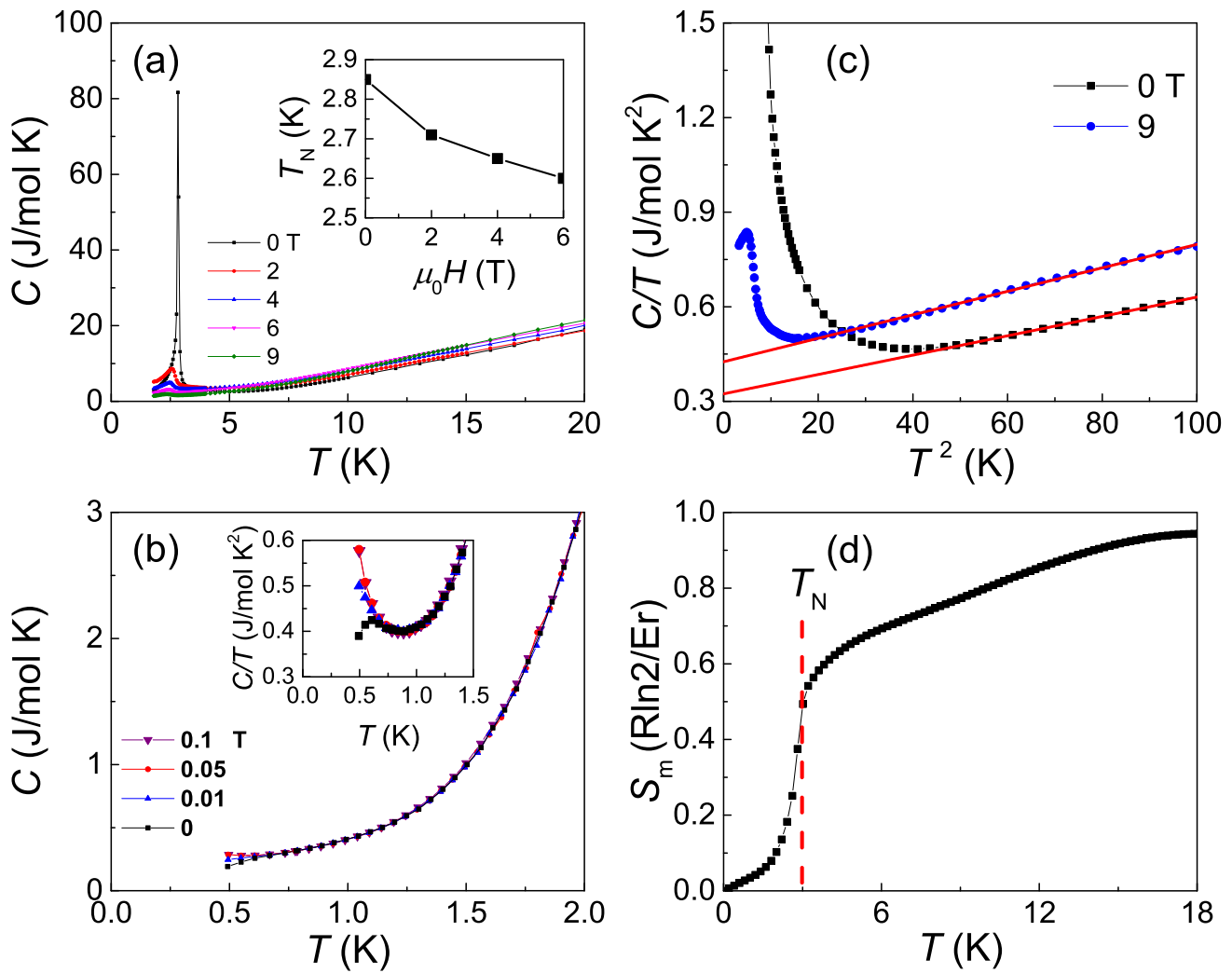

FIG. 3. (Color online) (a) Temperature dependence of the specific heat $C$ of $Er_2O_2Bi$ under magnetic fields $\mu_0H$ = 0, 2, 4, 6 and 9 T, respectively. Inset: magnetic field dependence of $T_N$ which is determined by the peak of $dC/dT$. (b) Temperature dependence of the specific heat $C$ of $Er_2O_2Bi$ among the temperature range 0 - 2 K under the magnetic fields $\mu_0H$ = 0, 0.01, 0.05 and 0.1 T, respectively. Inset: $C/T$ versus $T$ under different magnetic field $\mu_0H$ = 0, 0.01, 0.05 and 0.1 T, respectively. (c) Specific heat divided by temperature $C/T$ versus $T^2$ under magnetic fields $\mu_0H$ = 0 and 9 T, respectively. The red solid line represents the fitting by equation 2. (d) Magnetic entropy calculated by integrating $C/T$ (exclude phonon contribution) and $T$.

Figure 5(a) shows temperature dependence of resistivity of the samples $Er_2O_xBi$ with different oxygen content $x$. As $x$ increases, $T_c$ increases from 1 K to 1.23 K. We use the peak of $Td\chi/dT$ displayed in Figure 5(b) to define the $T_N$. In contrast to the SC transition, $T_N$ decreases with increasing $x$. The parameters, $n_H$, $T_c$ and $T_N$ of the samples are presented in the Figure 5(c). The increase in the carrier density can account for the increase of $T_c$ and the decrease of $T_N$, which reveals that the AFM order may compete with superconductivity.

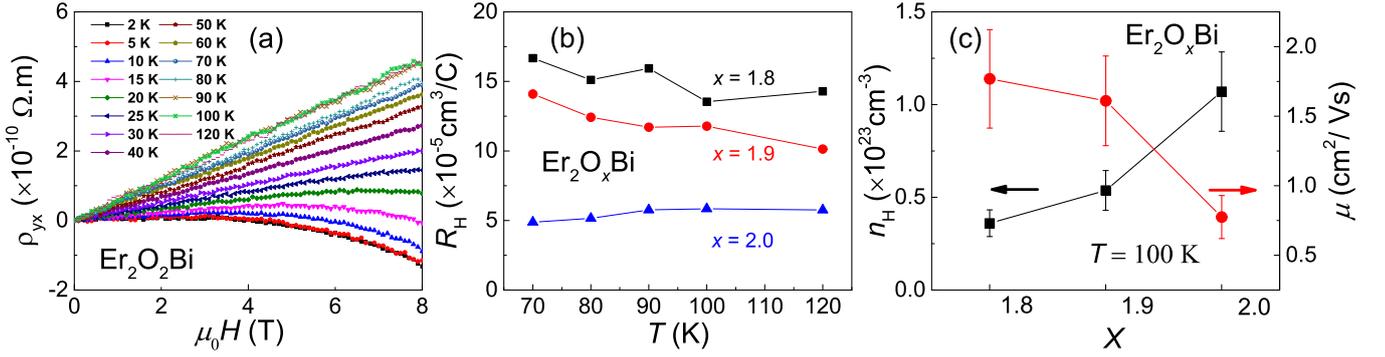

FIG. 4. (Color online) (a) Magnetic field dependence of Hall resistivity of the sample $Er_2O_2Bi$ at different temperatures. (b) Temperature dependence of the Hall coefficient $R_H$ of $Er_2O_xBi$ with $x = 1.8$, $1.9$ and $2.0$, respectively. The $R_H$ is determined from the initial slope of $\rho_{yx}(B)$ at $B \to 0$. (c) $x$ dependence of carrier density (left axis) and mobility (right axis) at 100 K. The connected lines are the guides to eyes.

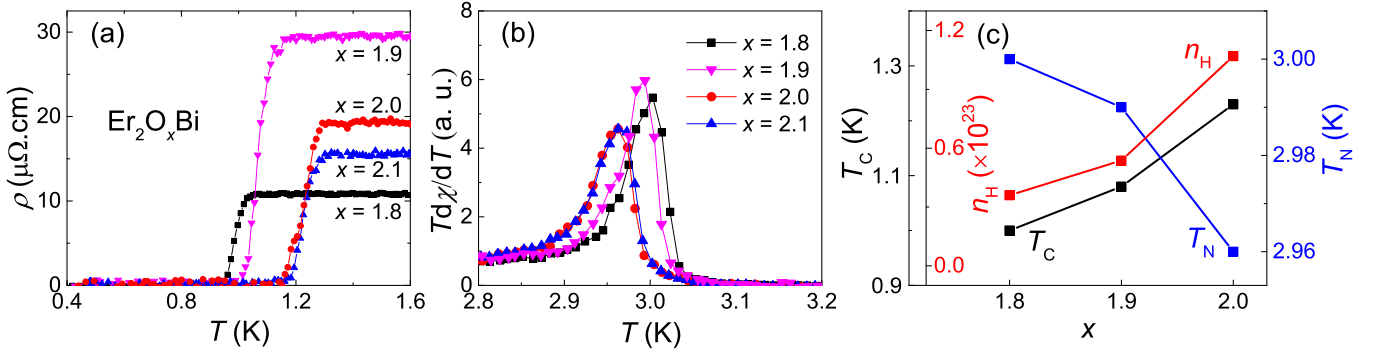

FIG. 5. (Color online) (a) Temperature dependence of resistivity of $Er_2O_xBi$. (b) Temperature dependence of the derivative susceptibility $Td\chi/dT$ of samples with different $x$. (c) $x$ dependence of $n_H$ (left axis) at $T = 100$ K, $T_c$ (left axis) and $T_N$ (right axis).

The primitive unit cell (with the simplest AFM configuration) and Brillouin zone of $Er_2O_2Bi$ are shown in Figure 6(a) and (b). The band structures of PM phase (black line) and AFM phase (red line) are shown in 6(c). In the AFM phase, the Er-$4f$ orbitals are away from the Fermi level, located around 6 eV (1 eV) below (above) Fermi energy $E_F$. Thus the electronic structure of AFM phase resembles that of PM phase near Fermi surface with two bands derived from Bi-$p$ orbits crossing $E_F$, which indicates that the magnetic moments of $Er^{3+}$ ions probably have little effect on the electronic structure near the Fermi level. This result may account for the similar electron-phonon coupling of $Er_2O_2Bi$ and $Y_2O_2Bi$. By integrating the $4f$ contribution below $E_F$, each Er ion has a spin polarization of 2.9 $\mu_B$ in the $z$ direction, corresponding to a spin configuration of $S = 3/2$. Therefore, the effective magnetic moment can be obtained with $\mu_{eff} = g_j \sqrt{(J(J+1))}\mu_B$. In this calculation, it is found $g_j = 1.2$, resulting in $\mu_{eff} = 9.58$ $\mu_B$, in good agreement with above experimental data, implying the fully localized nature of $4f$ electrons in $Er_2O_2Bi$ in the AFM ground state.

## IV. CONCLUSION

In summary, we have systematically investigated the superconducting properties and AFM order of the $Er_2O_2Bi$ compounds with different oxygen content by performing resistivity, magnetization, specific heat measurements and first-principle calculations. There may exist strong magnetic fluctuations or short-range AFM cor-relation above $T_N$ and they contributes to the significantly enhanced Sommerfeld coefficient of specific heat. The magnitude of $H_{c2}(0)$ and electron-phonon coupling constants in both $Er_2O_2Bi$ and $Y_2O_2Bi$ are very close, implying that superconductivity in both magnetic or nonmagnetic $R_2O_2Bi$ ($R$ = Y and Er) may share the same origin. The band structures from the first-principle calculations support the view that the charge carriers in $Er_2O_2Bi$ mainly come from the Bi square net layer. Therefore, the conducting monatomic $Bi^{2-}$ square net layer may be crucial to the occurrence of superconductivity in $R_2O_2Bi$. By adjusting the nominal oxygen content in $Er_2O_xBi$, we obtained several samples with different carrier density. The $Er_2O_2Bi$ compound is a typical example with the coexistence of superconductivity and AFM order in the series of $R_2O_2Bi$ ($R$ = rare earth or Y) materials. This study suggested that there may be a competition between superconductivity and AFM order in $Er_2O_2Bi$. It is worth further studying to understand the interplay of rare earth magnetism and superconductivity in such an anti-$ThCr_2Si_2$-type structure.

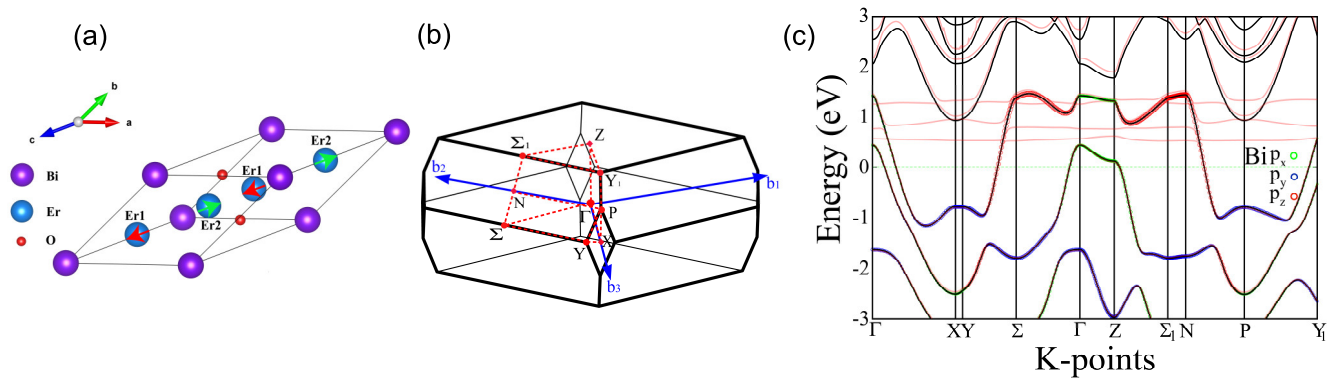

FIG. 6. (Color online) (a) Primitive unit cell (with the simplest AFM configuration) of $Er_2O_2Bi$. The $a$, $b$ and $c$ represent the directions of the lattice vectors. The directions of magnetic moments of different Er atoms in the simplest AFM configuration are also displayed. (b) Brillouin zone of $Er_2O_2Bi$. The $b_1$, $b_2$ and $b_3$ are the reciprocal lattice vectors and the red points are high symmetry points. (c) Band structures of paramagnetic phase (black line) and AFM phase (red line). The hollow circles with different colours represents the different orbits of Bi.

## ACKNOWLEDGMENTS

The authors would like to thank Yifeng Yang, and Jianhui Dai for insightful discussions. This work was supported by the National Key R&D Projects of China (Grant Nos. 2019YFA0308602 and 2016YFA0300402), the National Science Foundation of China (Grant No. 11774305), and the Fundamental Research Funds for the Central Universities of China.

---

* zhuan@zju.edu.cn